\documentclass[journal=jpcafh, manscript=article, layout=twocolumn, amsmath,amssymb,nobibnotes,unsortedaddress,superscriptaddress,usetitle=true]{achemso}

\setkeys{acs}{usetitle=true}
\usepackage{achemso}
\usepackage{graphicx}% Include figure files
\usepackage{tabularx}
\usepackage{dcolumn}% Align table columns on decimal point
\usepackage{bm}% bold math
\usepackage{multirow}
\usepackage{amsmath}
\usepackage{amssymb}
\usepackage{txfonts}
\usepackage{url}
\usepackage[usenames,dvipsnames]{color}
\definecolor{Gray}{gray}{0.0}
\definecolor{lightGray}{gray}{0.35}
\renewcommand{\deg}{$^\circ$}

\SectionNumbersOff

\title{Diffusion Monte Carlo evaluation of disiloxane linearisation barrier}% Force line breaks with \\

\author{Adie Tri Hanindriyo$^{1,*}$}
\email{adietri@icloud.com}
\author{Amit Kumar Singh Yadav$^{2}$}
\author{Tom Ichibha$^{3}$}
\author{Ryo Maezono$^{4}$}
\author{Kousuke Nakano$^{4,5}$}
\author{Kenta Hongo$^{6}$}

\affiliation{$^{1}$
  School of Materials Science, JAIST, Asahidai 1-1, Nomi, Ishikawa,
  923-1292, Japan
}

\affiliation{$^{2}$
  Department of Electrical Engineering, Indian Institute of Technology Gandhinagar,
  Palaj 382355, Gujarat, India
}

\affiliation{$^{3}$
  Materials Science and Technology Division, 
  Oak Ridge National Laboratory, Oak Ridge, Tennessee 37831, USA
}

\affiliation{$^{4}$
  School of Information Science, JAIST, Asahidai 1-1, Nomi, Ishikawa,
  923-1292, Japan
}

\affiliation{$^{5}$
  Scuola Internazionale Superiore di Studi Avanzati(SISSA),
  via Bonomea, 265-34136 Trieste ITALY.
}

\affiliation{$^{6}$
  Research Center for Advanced Computing Infrastructure, JAIST,
  Asahidai 1-1, Nomi, Ishikawa 923-1292, Japan
}

\begin{document}
\begin{abstract}
\noindent
The disiloxane molecule is a prime example of silicate compounds
containing the Si-O-Si bridge. The molecule is of significant interest within the field of quantum chemistry,
owing to the difficulty in theoretically predicting its properties.
Herein, the linearisation barrier of disiloxane is investigated using
a fixed-node diffusion Monte Carlo (FNDMC) approach,
which is currently the most reliable
{\it ab initio} method in accounting for an electronic correlation.
Calculations utilizing the density functional theory (DFT) and the
coupled cluster method with single and double substitutions, 
including noniterative triples (CCSD(T))are carried out alongside FNDMC
for comparison.
Two families of basis sets are used
to investigate the disiloxane linearisation barrier - Dunning's
correlation-consistent basis sets cc-pV$x$Z ($x = $ D, T, and Q)
and their core-valence correlated counterparts, cc-pCV$x$Z.
It is concluded that FNDMC successfully predicts
the disiloxane linearisation barrier and does not depend
on the completeness of the basis sets as much as DFT or CCSD(T), thus establishing its suitability.
%\\
%% {\it Keywords: {\it ab initio}; quantum Monte Carlo; silicate; disiloxane; linearisation barrier}
\end{abstract}
%------------------

\maketitle

%\tableofcontents

%%%%%%%%%%%%%%%%%%%%%%%%%%%%%%%%%
\section{\label{sec:intro}INTRODUCTION}
%%%%%%%%%%%%%%%%%%%%%%%%%%%%%%%%%
The simplest molecule containing the Si-O-Si bond is disiloxane or Si$_2$H$_6$O.
Also called disilyl ether, its structure is a single Si-O-Si bond terminated
by three H atoms at each end (H$_3$Si-O-SiH$_3$).
There have been numerous studies
investigating the Si-O-Si bond,
particularly owing to its importance
in the modelling of silica compounds, which
are the most abundant constituent of the Earth's crust.
Most importantly, silica compounds range in function from glasses to quartz crystals,
both of which comprise large sectors in industry.
In some studies, disiloxane has been used as a sealant and as a component in cosmetics,
or as a prototype region of a zeolite or clay substrate for applications ranging from catalysis to prebiotic synthesis~\cite{1993LUK}.

\vspace{2mm}
Experimental evaluations of the Si-O-Si bond indicate
an anharmonic bending potential with a low linearisation barrier,
which makes it quite difficult to attain sufficient accuracy
in such measurements~\cite{1977DUR,1983KOP}.
Despite the significant volume of previous studies
dedicated to the Si-O-Si
bond~\cite{1990KOP,1993LUK,1994CSO,1995KOP,2007CAR,2008DER,2015NOR,2019NOR},
the properties of Si-O-Si bond obtained in most of these studies are not consistent with each other. 
In 
{\it ab initio} studies, in particular, multiple calculation methods have resulted in
different values for the bond angle and length~\cite{1995NIC,2003YUA,2015NOR}, as well as the
linearisation energy~\cite{1990KOP,1994CSO,2008DER} and
Si-O-Si potential energy surface~\cite{1993LUK,2015NOR}, among other factors
These properties and the bond geometry itself are shown
to be sensitive to the choice of basis set and
level of theory (the former more than the latter)
according to at least one previous study~\cite{2008DER}.

\vspace{2mm}
To narrow down the possibilities,
it would be ideal to employ the most
reliable methods at our disposal.
Quantum Monte Carlo (QMC),
currently the 
most reliable of the many-body calculation
methods, is expected to provide a reasonable
and reliable prediction.~\cite{2001FOU}
We used the
fixed-node diffusion Monte Carlo (FNDMC) 
method, which has been widely and successfully applied 
to several molecular systems~\cite{2010HON,2012WAT,2013HON,2015HON,2008KOS,2017HON,2019TAK,2017ICH}.
Although FNDMC results are also affected 
by the choice of basis sets~\cite{2008KOS,2019NAK},
we note that the basis-set dependence
is considerably different from that for 
an SCF-based method, such as the density functional theory (DFT) 
and molecular orbital (MO) methods.
In SCF-based approaches, 
the choice of basis set 
affects both the amplitude 
and the nodal positions 
of the corresponding 
many-body wavefunctions 
(although the methods do not explicitly
employ a many-body wavefunction condition). 
With FNDMC, by contrast, 
the choice only affects 
the nodal positions. 
The amplitude can be automatically 
adjusted such that its shape 
may approach that of 
the exact solution as closely as possible under a restriction 
with a fixed nodal position~\cite{2001FOU,2009MAE}.
A typical example is the description 
of electron-nucleus cusps.~\cite{2005MA}
Even when using such a poor basis set that cannot describe 
the singularity of the cusp 
based on its analytical form, 
an initial guess is applied
for further numerical evolution 
driven by the FNDMC, making the 
amplitude at the nucleus positions 
singular with a cusp~\cite{2001FOU,2009MAE}.
Owing to this self-healing property 
of the amplitude, 
the dependence on the choice of 
basis sets in FNDMC 
becomes considerably weaker. That is, the bias arising from the choice 
is reduced more than 
that for SCF-based ones, 
which is a difficulty faced by the 
present systems.~\cite{2008DER}
Note that ``cuspless'' basis sets cause
a singularity of the local energy in FNDMC, thus resulting in numerical instability.
However, this singularity can be easily remedied by
introducing the cusp correction
proposed by Ma et al~\cite{2005MA}.

\vspace{2mm}
In this study, we applied FNDMC to
investigate the basis-set dependence of
the linearization energy of a disiloxane molecule in
comparison with other {\it ab initio} methods including DFT and CCSD(T), as well as
empirical measurements from earlier studies conducted on disiloxane~\cite{1960ARO,1977DUR,1983KOP}.
The FNDMC depends on the basis set implicitly through its fixed nodal surface
provided by the Slater determinant whose orbitals are expanded
in terms of the basis set,
whereas DFT and CCSD(T) strongly depend on the choice of basis set.
The cc-pV$x$Z~\cite{1989DUN,1992KEN,1993WOO} and
cc-pCV$x$Z~\cite{2002PET} basis sets ($x =$ D, T, Q), which are
standard correlation methods,
were examined in the present study and then applied
to all the {\it ab initio} calculations
to investigate the basis-set dependence of linearisation energy evaluation.

%%%%%%%%%%%%%%%%%%%%%%%%%%%%%%%%%
\section{\label{sec:methods}MODEL AND METHODOLOGY}
\begin{figure}[htbp]
  \centering
    \includegraphics[width=\linewidth]{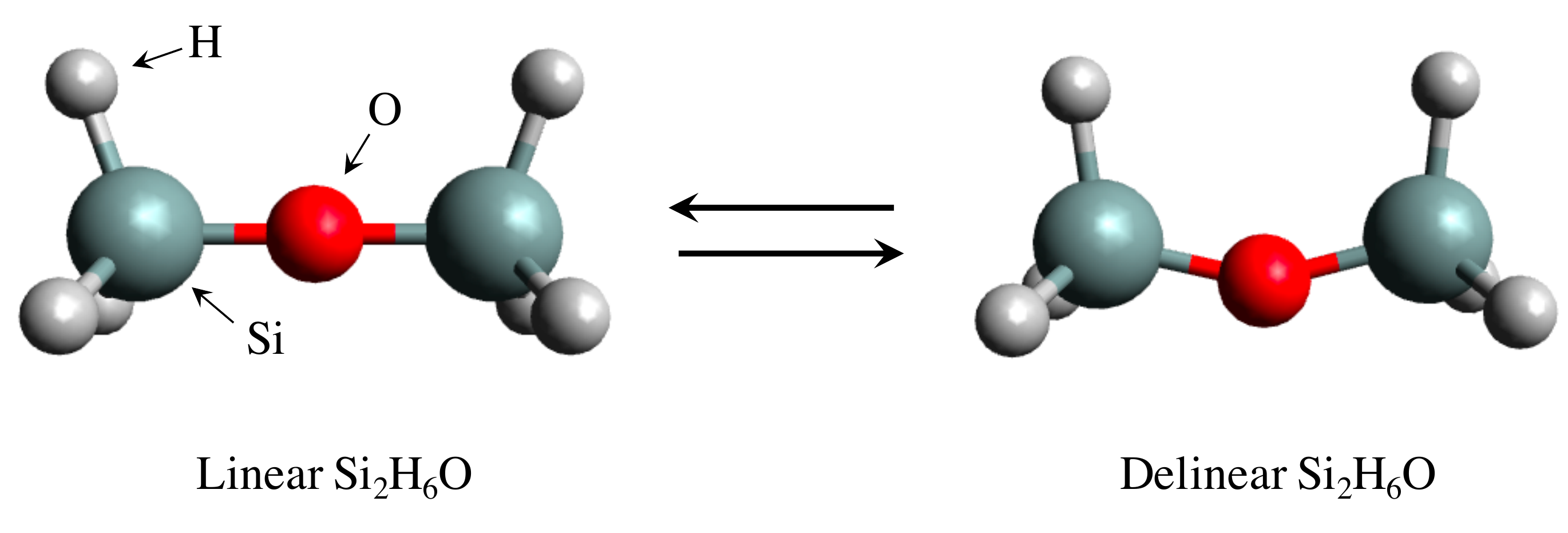}
    \caption{Linear and non-linear molecular structures of disiloxane, Si$_2$H$_6$O.}
    \label{fig:struc}
\end{figure}

Our target property is simply a linearisation barrier of disiloxane Si$_2$H$_6$O
between ``linear'' and ``non-linear'' (bent) structures.
As can be seen from Figure~\ref{fig:struc},
the linear structure possesses an eclipsed ${\rm D_{3h}}$ symmetry~\cite{2007CAR}
and a non-linear structure is a bent conformation
of the ${\rm C_{2v}}$ symmetry~\cite{1963ALM}.
For the two fixed structures,
the barrier, $\Delta E_{\rm{barrier}}$, is
defined as the energy difference
between the linear and non-linear structures:
%%%%%%%%%%%%%%%%%%%%%%%%%%%%%%%%%%%%%%%%%%%%%%%%%%%%%%%%%%%%%%%%%%%%%%%%%%%%%%%%%%%
\begin{eqnarray}\label{eq:LinBar}
  \Delta E_{\rm barrier} = {E_{\rm linear}} - {E_{\rm delinear}},
\end{eqnarray}
%%%%%%%%%%%%%%%%%%%%%%%%%%%%%%%%%%%%%%%%%%%%%%%%%%%%%%%%%%%%%%%%%%%%%%%%%%%%%%%%%%%
where ${E_{\rm linear}}$ and ${E_{\rm delinear}}$ are the (electronic) total energies of 
the linear and non-linear structures, respectively.
%FFFFFFFFFFFFFFFFFFFFFFFFFFFFFFFFFFFFFFFFFFFFFFFFFFFFFFFFFFFFFFFFFFF
%FFFFFFFFFFFFFFFFFFFFFFFFFFFFFFFFFFFFFFFFFFFFFFFFFFFFFFFFFFFFFFFFFFF

\vspace{2mm}
The accuracy of computed $\Delta E_{\rm{barrier}}$ values can be calibrated
by referring to experimentally observed $\Delta E_{\rm{barrier}}$ values.
To compare our computational results with the experimental results, however,
we note that the experimental $\Delta E_{\rm{barrier}}$ involves
the zero-point energy (ZPE) contribution, 
whereas our computational $\Delta E_{\rm{barrier}}$ value
considers only the electronic contribution.
The ZPE contributions can also be evaluated at each level of theory,
although their accuracy depends on methods adopted and the basis sets.
To investigate the basis-set dependence of $\Delta E_{\rm{barrier}}$,
we simply consider the electronic $\Delta E_{\rm{barrier}}$.
Thus, the experimental $\Delta E_{\rm{barrier}}$ values are
corrected by adopting a theoretical estimate from Koput~\cite{1995KOP}.
We address this issue in more detail later.

\vspace{2mm}
In the present study, the above two energies
${E_{\rm linear}}$ and ${E_{\rm delinear}}$ were computed through
various combinations of (1) {\it ab initio} methods and (2) basis sets

\vspace{2mm}
(1) Our methods applied are 
DFT with B3LYP exchange-correlation functional~\cite{1993BEC} (DFT-B3LYP),
CCSD(T)~\cite{1977MON},
and FNDMC~\cite{2001FOU} with trial wavefunctions generated from DFT-B3LYP.
Within the framework of DFT, DFT-B3LYP is a standard method
for covalent systems. Within the correlated methods,
CCSD(T) is known to be the ``gold standard'' in quantum chemistry.
A variety of methods and basis sets were chosen
in line with the previous results~\cite{2008DER} indicating
dependence of the Si-O-Si bond description on such choice.
FNDMC is known to be comparable with CCSD(T);
however, for the first time,
we applied FNDMC to an evaluation of $\Delta E_{\rm{barrier}}$.

\vspace{2mm}
(2) Our basis sets are 
a family of Dunning's correlation-consistent basis sets
(cc-pV$x$Z; $x$ = D, T, Q)~\cite{1989DUN,1992KEN,1993WOO}
and their core-valence correlated variants
(cc-pCV$x$Z; $x$ = D, T, Q)~\cite{2002PET}.
The cc-pV$x$Z and cc-pCV$x$Z basis sets were originally
developed for correlated methods such as CCSD(T),
but have been found to be
appropriate even for DFT-B3LYP and FNDMC
in properly reproducing the dynamic electron correlation.~\cite{1989DUN}
In particular, the polarisation functions included implicitly in cc-pV$x$Z
are essential for properly describing the Si-O-Si bond.
To describe the correlation effects more precisely, 
the present study considers the dynamic correlation
between core and valence electrons by applying the cc-pCV$x$Z basis sets,
which can reproduce the core-valence correlation 
by minimising the difference in the correlation energies
between all-electron and valence-only (using pseudopotentials)
calculations.~\cite{1995WOO}
The number of basis functions entailed in each basis set
is shown in Table~\ref{table:bs}.
As shown, accounting for the core-valence correlation is
more expensive for increasingly complete orbital and polarisation functions,
with a quadruple zeta-level basis set
adding close to 130 more basis functions
to the standard correlation-consistent basis set.
The number of basis functions is expected
to largely correlate with the reliability of the calculations,
particularly for the core-valence variants pertaining to all-electron calculations.
%TTTTTTTTTTTTTTTTTTTTTTTTTTTTTTTTTTTTTTTTTTTTTTTTTTTTTTTTTTTTTTTTTTTT
\begin{table}[htbp]
 \centering
 \caption{List of basis functions}
 \label{table:bs}
 %\resizebox{\textwidth}{15mm}{
 \begin{tabular}{lc}
  \hline
  Basis set & Linear basis functions\\
  \hline
  cc-pVDZ         & \phantom{0}80  \\
  cc-pCVDZ        &  102  \\
  cc-pVTZ         &  182  \\
  cc-pCVTZ        &  245  \\
  cc-pVQZ         &  353  \\
  cc-pCVQZ        &  482  \\
  \hline  
 \end{tabular}
\end{table}
%TTTTTTTTTTTTTTTTTTTTTTTTTTTTTTTTTTTTTTTTTTTTTTTTTTTTTTTTTTTTTTTTTTTT%

\vspace{2mm}
In the present study we obtained the linear and delinear structures
using the second-order M\o{}ller-Plesset perturbation theory (MP2)
with the cc-pVQZ basis set (MP2/cc-pVQZ level),
and previous studies conducted at the MP2/cc-pVQZ level of theory~\cite{1995NIC,2007CAR}
have established a reliable
favourability toward a delinear structure and
provide good agreement for the Si-O-Si angle
in accordance with experimental results.~\cite{1963ALM}
Modelling of the Si-O-Si angle in particular has been heavily emphasised
in previous theoretical studies for both disiloxane~\cite{2008DER}
and pyrosilisic acid~\cite{2015NOR,2019NOR}.
This necessitates even larger numbers of basis functions,
which means that the cc-pCVQZ basis set would necessitate
the highest number of basis functions by far.
The two optimised structures at the MP2/cc-pVQZ level of theory 
were used in this work to
calculate all $\Delta E_{\rm{barrier}}$ values,
i.e., common to all levels of theory, 
where the linear structure has a Si-O-Si angle of 180\deg\
and the delinear structure has an optimised Si-O-Si angle of 146.8\deg\
comparable with the experimental value of 144.1\deg~\cite{1963ALM}.
The other structural parameters are given in the Supporting Information,
and are also in good agreement with experiments.

\vspace{2mm}
The present study adopted no pseudopotential calculations,
but rather all-electron calculations 
with a total of 42 electrons for a single disiloxane molecule.
This molecular system is not so large that the CCSD(T) calculation
within the frozen core approximation is feasible,
despite the computational cost of CCSD(T) scaling as $N^7$,
where $N$ is the number of electrons in the system.
By contrast, the DFT cost scales with $N^3$ (or less) and is therefore unimportant.
Similar to DFT, the FNDMC cost scales as $N^3$,
although the prefactor of FNDMC is much larger than that of DFT.
This is because a vast number of random sampling points are required to
obtain a sufficiently small error bar (sub-chemical accuracy of $\sim 0.1$ kcal/mol)
to calibrate a small $\Delta E_{\rm{barrier}}$ ($\sim 0.5$ kcal/mol).
Recent parallel computers, however, enable us to apply FNDMC
to evaluate such a tiny $\Delta E_{\rm{barrier}}$,
because its algorithm is intrinsically parallel.

\vspace{2mm}
In our FNDMC calculations, we adopted Slater-Jastrow type wavefunctions
as their fixed-node trial wavefunctions~\cite{1955JAS}.
Molecular orbitals entering the single Slater determinants were
generated by DFT-B3LYP with various types of basis sets.
The Jastrow factor consists of one-, two-, and three-body
terms~\cite{2004DRU} including 88 variational parameters in total.
These parameters were optimised through a variance minimisation scheme~\cite{2005DRU}, and
only the two-body term holds the electron-electron cusp condition~\cite{1957KAT}.
The electron-nucleus cusp condition, which is a short-range one-body correlation effect, 
is satisfied by the cusp correction scheme applied to the Gaussian basis sets~\cite{2005MA}
instead of imposing the cusp condition on the one-body term.
Note that the Jastrow factor does not change the (fixed) nodal surfaces
and is responsible for the numerical stability in FNDMC.
However, the quality of the nodal surfaces determines
the accuracy of the FNDMC energies
in terms of the fixed-node variational principles~\cite{1982REY}.
Accordingly, the FNDMC accuracy depends implicitly on the basis set adopted,
which is used to expand the molecular orbitals entering the Slater determinant.
In addition to the fixed-node error, another source of bias
in actual FNDMC calculations arises from the short-time approximation
with finite (small) timesteps~\cite{1993UMR}.
To remove this bias, it is common to use multiple timesteps
to make the linear regression and obtain
calculation results for $\delta t \rightarrow 0$.
This regression is an approximation of the theoretical $\delta t = 0$ result.
The present study considers both linear and quadratic extrapolations.

\vspace{2mm}
The software package Gaussian09~\cite{2009GAU}
was used to conduct the MP2 (geometry optimisation),
DFT-B3LYP, and CCSD(T) calculations,
whereas the CASINO~\cite{2010CAS} code was used
to apply the FNDMC calculations.
An electron-nucleus cusp correction scheme~\cite{2005MA}
for Gaussian orbitals was utilised
in the all-electron FNDMC calculation in CASINO.
In addition, cc-pCV$x$Z ($x$ = D, T, Q) basis sets for Si and O
as well as the corresponding cc-pV$x$Z versions for
the H atoms during the same calculations
were obtained from the online Basis Set Exchange library~\cite{2019PRI}.
In FNDMC for both structures,
the number of target population was 11,520, and 
the number of steps in the imaginary-time evolution
was set to 2,000 and 500,000 for equilibrated and
accumulated phases, respectively. In addition, we carried out FNDMC calculations
with different timesteps of 0.01, 0.005, and 0.001
to remove the short-time bias (see SI for more details).

%%%%%%%%%%%%%%%%%%%%%%%%%%%%%%%%%
\section{\label{sec:resultsdiscussion}RESULTS and DISCUSSION}
%%%%%%%%%%%%%%%%%%%%%%%%%%%%%%%%%%%%%%%%%%%%%%%%%%%%%%%%%%%%%%%%%%%%%%%%%%%%%%%%%%%
\subsection{Geometry and ZPE contribution}
%%%%%%%%%%%%%%%%%%%%%%%%%%%%%%%%%%%%%%%%%%%%%%%%%%%%%%%%%%%%%%%%%%%%%%%%%%%%%%%%%%%
Several studies have been conducted on
the geometry of the disiloxane molecule,
although most {\it ab initio} methods tested did not manage
to replicate the available experimental results.
Rather than the Si-O bond length,
the Si-O-Si bond angle and linearisation barrier have been found
to be greatly dependent on the choice of basis set
used to represent the wavefunction~\cite{2008DER}.
This study focuses on the linearisation barrier of disiloxane,
taking the same optimised geometries for all calculations,
partly owing to the notorious difficulty in optimising the geometries in FNDMC.
%TTTTTTTTTTTTTTTTTTTTTTTTTTTTTTTTTTTTTTTTTTTTTTTTTTTTTTTTTTTTTTTTTTTTTTTTTTTTTTTT
\begin{table*}[htbp]
 \centering
 \caption{Linearisation barrier values obtained
   from theoretical calculations and experiments.
   Its ZPE correction and ZPE-corrected Raman value
   are also given.
   Computational values ($\Delta \epsilon_0$)
   are to be compared with a ZPE-corrected experimental value
   ($\Delta \epsilon_1 - \Delta \mathrm{ZPE}$);
   see the text for the definition of notation, sign, {\it etc. }.}
 \label{table:all}
 \begin{tabular}{lcccc}
  \multirow{2}{*}{Basis set} & \multicolumn{4}{c}{$\Delta \epsilon_0$ [kcal/mol]}\\\cline{2-5}
  & DFT-B3LYP & CCSD(T) & & FNDMC\\
  \hline
  cc-pVDZ         & \phantom{-}0.80 & \phantom{-}1.44 & &  \phantom{-}0.40 $\pm$ 0.15\\
  cc-pCVDZ        & \phantom{-}0.76 & \phantom{-}1.42 & &  \phantom{-}0.54 $\pm$ 0.12\\
  cc-pVTZ         & \phantom{-}0.18 & \phantom{-}0.46 & &  \phantom{-}0.47 $\pm$ 0.14\\
  cc-pCVTZ        &           -0.01 & \phantom{-}0.32 & &  \phantom{-}0.31 $\pm$ 0.11\\
  cc-pVQZ         & \phantom{-}0.18 & \phantom{-}0.44 & &  \phantom{-}0.26 $\pm$ 0.10\\
  cc-pCVQZ        & \phantom{-}0.06 & \phantom{-}0.35 & &  \phantom{-}0.36 $\pm$ 0.10\\
  \hline
  \hline
  \multicolumn{3}{c}{$\Delta \epsilon_1$ [kcal/mol]} & \multicolumn{1}{c}{} & \multicolumn{1}{c}{$\Delta \epsilon_1-\Delta$ZPE [kcal/mol]} \\ \cline{1-3} \cline{5-5}
  Far IR spectrum~\cite{1960ARO} & IR-Raman (solid)~\cite{1977DUR} & Raman~\cite{1983KOP} & & ZPE-corrected Raman ($\Delta$ZPE)~\cite{1995KOP}\\
  \multicolumn{1}{c}{1.1-1.4} & 0.32 & 0.3 & & 0.36 (-0.06) \\
  \hline
 \end{tabular}
\end{table*}
%TTTTTTTTTTTTTTTTTTTTTTTTTTTTTTTTTTTTTTTTTTTTTTTTTTTTTTTTTTTTTTTTTTTTTTTTTTTTTTTT

\begin{figure}[htbp]
  \centering
    \includegraphics[width=\linewidth]{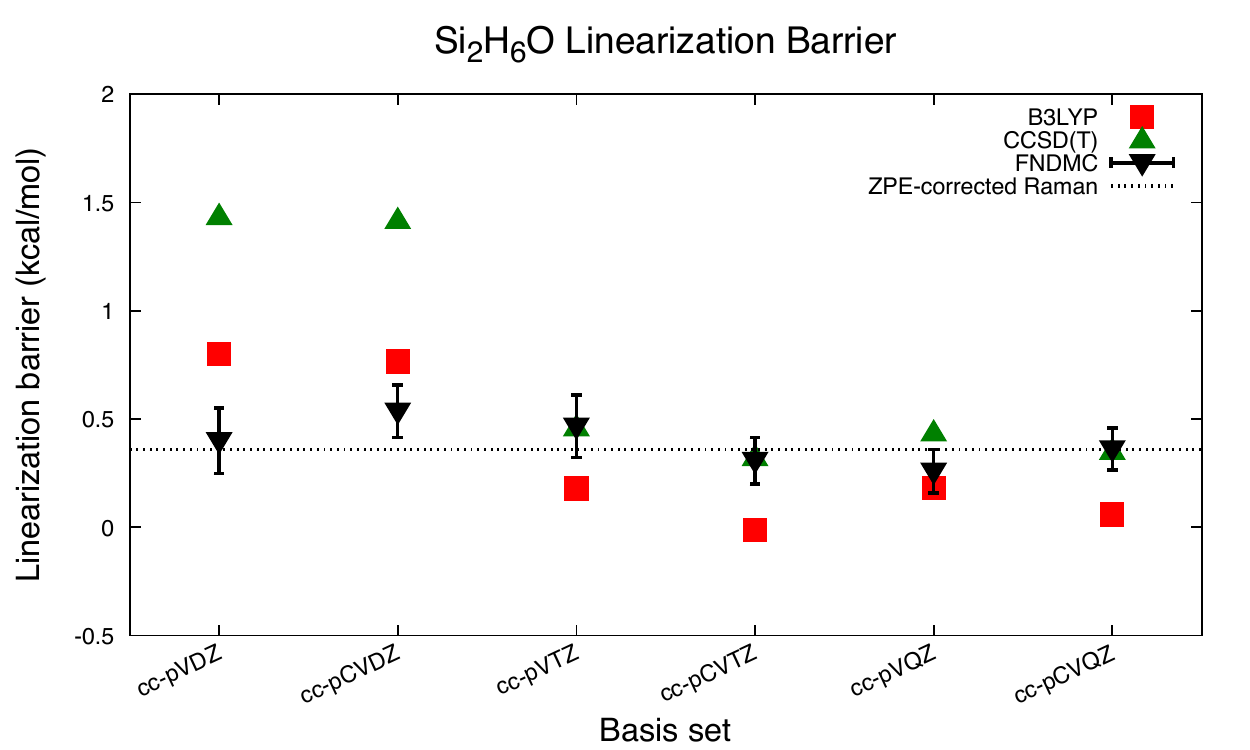}
    \caption{Linearisation barrier of disiloxane calculated in this work.}
    \label{fig:all}
\end{figure}

\vspace{2mm}
The computational and experimental 
linearisation barriers of disiloxane, $\Delta E_{\rm{barrier}}$,
defined in Eq.~\eqref{eq:LinBar}
are listed in Table~\ref{table:all} and plotted in Figure~\ref{fig:all}
as a visual guide.
It has been commonly observed experimentally
that the non-linear (bent) structure of disiloxane
is energetically more favourable than a linear structure,
which translates to a positive linearisation barrier.
All three experimental results seem to agree on this point,
and~\cite{1960ARO} in particular reports
a higher linearisation barrier (at 1.1 to 1.4 kcal/mol)
than the other two results~\cite{1977DUR,1983KOP},
which report a barrier of approximately 0.3 kcal/mol.
Both of these measurements are more recent than the first
and achieve good consilience with the DFT predictions
from the highest quality basis sets~\cite{2008DER}.
Therefore, it is reasonable to infer that a linearisation barrier
of approximately 0.3 kcal/mol is a reliable value for disiloxane.
%%%%%%%%%%%%%%%%%%%%%%%%%%%%%%%%%%%%%%%%%%%%%%%%%%%%%%%%%%%%%%%%%%%%%%%%%%%%%%%%%%
%%%%%%%%%%%%%%%%%%%%%%%%%%%%%%%%%%%%%%%%%%%%%%%%%%%%%%%%%%%%%%%%%%%%%%%%%%%%%%%%%%

\vspace{2mm}
It needs to be stressed, however,
that the experimentally measured linearisation barrier
may not be comparable to the ground state values calculated from the first principles.
The ground state energy obtained through {\it ab initio} calculations
are physically unobtainable experimentally because of the ZPE, i.e. 
the difference between the ground state and the lowest energy vibrational state.
Even at absolute zero temperature,
the lowest energy level achievable is a vibrational state $\epsilon_1$
instead of the electronic ground state $\epsilon_0$:
%EEEEEEEEEEEEEEEEEEEEEEEEEEEEEEEEEEEEEEEEEEEEEEEEEEEEEEEEEEEEEEEE
\begin{eqnarray}\label{eq:ZPE}
  \epsilon_1 = \epsilon_0 + {\rm ZPE}
\end{eqnarray}
%EEEEEEEEEEEEEEEEEEEEEEEEEEEEEEEEEEEEEEEEEEEEEEEEEEEEEEEEEEEEEEEE
Consequentially, any energetic barriers measured in the experiments
is at best the difference
between the lowest vibrational states $\Delta\epsilon_1$,
whereas energetic barriers calculated through {\it ab initio} methods
are from the electronic ground states $\Delta\epsilon_0$.
Therefore, a comparison between energetic barriers obtained
from the theoretical calculations and experimental measurements
must also account for the difference in ZPE.
The energetic barrier of a transition
from quantum state A to state B is calculated as follows:
%EEEEEEEEEEEEEEEEEEEEEEEEEEEEEEEEEEEEEEEEEEEEEEEEEEEEEEEEEEEEEEEE
\begin{eqnarray}\label{eq:Delta}
  \epsilon^{\rm B}_1-\epsilon^{\rm A}_1 &=& (\epsilon^{\rm B}_0 + {\rm ZPE}^{\rm B}) - (\epsilon^{\rm A}_0 + {\rm ZPE}^{\rm A})\nonumber\\
  \epsilon^{\rm B}_1-\epsilon^{\rm A}_1 &=& (\epsilon^{\rm B}_0 - \epsilon^{\rm A}_0) + ({\rm ZPE}^{\rm B} - {\rm ZPE}^{\rm A})\nonumber\\
  \Delta\epsilon_1 &=& \Delta\epsilon_0 + \Delta{\rm ZPE}\nonumber\\
  \Delta\epsilon_0 &=& \Delta\epsilon_1 - \Delta{\rm ZPE}
\end{eqnarray}
%EEEEEEEEEEEEEEEEEEEEEEEEEEEEEEEEEEEEEEEEEEEEEEEEEEEEEEEEEEEEEEE
The difference in ZPE $\Delta{\rm ZPE}$
between the initial and final states A $\to$ B is
usually considered insignificant.
However, barrier height discrepancies on the order of 0.1 kcal/mol
might well be caused by this term.

\vspace{2mm}
The preceding theoretical study by Koput~\cite{1995KOP}
estimates a $\Delta$ZPE value of $-20$ cm$^{-1}\approx -0.06$ kcal/mol,
which in line with Equation~\ref{eq:Delta}
should result in a higher ground state linearisation barrier
than the experimentally measured result.
Therefore, this study treats the ground state
linearisation barrier of 0.36 kcal/mol
as a reasonably accurate ``exact" linearisation barrier
for a point of comparison with {\it ab initio} calculations.
This value is referred to as 
``ZPE-corrected Raman'' in Table~\ref{table:all}
and based on a dotted line in the figures herein, as a point of comparison.
ZPE correction has not been considered
in {\it ab initio} comparisons or based on experimental results,
which have cited 0.3 kcal/mol as the point of comparison.~\cite{2008DER}

%%%%%%%%%%%%%%%%%%%%%%%%%%%%%%%%%%%%%%%%%%%%%%%%%%%%%%%%%%%%%%%%%%%%%%%%%%%%%%%%
\subsection{DFT-B3LYP results}
%%%%%%%%%%%%%%%%%%%%%%%%%%%%%%%%%%%%%%%%%%%%%%%%%%%%%%%%%%%%%%%%%%%%%%%%%%%%%%%%
Both the cc-pV$x$Z and cc-pCV$x$Z basis sets
show positive linearisation barriers (except for cc-pCVTZ),
energetically favouring the non-linear over the linear conformer of disiloxane.
The simplest basis set, i.e. the double zeta cc-pVDZ basis set,
results in the highest barrier of all (0.80 kcal/mol),
in certain agreement with the far infrared absorption spectra
of gaseous disiloxane~\cite{1960ARO}.
The core-valence correlated counterpart, i.e.
the cc-pCVDZ basis set, also produces a high linearisation barrier of 0.78 kcal/mol. 
Triple zeta basis sets (cc-pVTZ and cc-pCVTZ) show
a significant reduction in the barrier height,
eventually converging to the quadruple zeta set results.
These results reinforce the conclusions drawn
in preceding studies~\cite{1994CSO,1995KOP,1995NIC,2007CAR,2008DER}
that the basis sets at the double zeta level are insufficient
to properly describe the disiloxane molecular properties,
and that the triple zeta level is the minimally reliable level
for the description of the atomic orbitals.
The trends of the linearisation barrier, meanwhile,
is nearly identical for both standard and core-correlated variants
(Figure~\ref{fig:cc}), with higher quality sets producing
good agreement with two of the three available experimental
results~\cite{1977DUR,1983KOP}.

\vspace{2mm}
The results for DFT-B3LYP in this study,
particularly the converged results for
the quadruple zeta basis sets cc-pVQZ and cc-pCVQZ,
seem to indicate that previous agreement
between the B3LYP results and experimental data~\cite{2008DER}
is due to coincidence instead of convergence
toward the complete basis set limit.
Indeed, for the largest basis sets used in this study,
the linearisation barrier calculated by B3LYP
is at best approximately 0.1 kcal/mol smaller than the ZPE-corrected measurement.
It does not seem that the B3LYP exchange-correlation functional
provides adequate accounting of the electron correlation
for properly describing a disiloxane molecule.

\subsection{CCSD(T) results}
CCSD(T) calculations were conducted with the same geometries and basis sets as the B3LYP calculations.
The results of these calculations are shown in Table~\ref{table:all}.
The trends of the linearisation barrier agree almost perfectly with B3LYP calculations,
with a universal shift in the energetic favourability
toward the non-linear conformer denoted by the larger linearisation barriers.
CCSD(T) calculations seem to converge to
values closer to the ZPE-corrected Raman linearisation barrier of 0.36 kcal/mol.
The cc-pV$x$Z basis sets converge to a linearisation barrier of approximately 0.44 kcal/mol,
whereas the core-valence correlated set cc-pCV$x$Z converges to a barrier height of approximately 0.35 kcal/mol,
in excellent agreement with the benchmark value.
We also achieved good agreement with another CCSD(T) calculation
in an earlier study~\cite{2008DER} reporting a linearisation barrier of 0.48 kcal/mol using the cc-pVTZ basis set.
These results show that accounting for the electron correlation is
indeed necessary to properly model the Si-O-Si bond in disiloxane,
and supports the conclusion of earlier studies citing cc-pVTZ and CCSD(T)
as the minimum reliable level of description for the disiloxane molecule~\cite{2007CAR,2008DER}.

%%%%%%%%%%%%%%%%%%%%%%%%%%%%%%%%%%%%%%%%%%%%%%%%%%%%%%%%%%%%%%%%%%%%%%%%%%%%%%%%
\subsection{FNDMC results}
%%%%%%%%%%%%%%%%%%%%%%%%%%%%%%%%%%%%%%%%%%%%%%%%%%%%%%%%%%%%%%%%%%%%%%%%%%%%%%%%
The linearisation barrier calculated by FNDMC is also shown in Table~\ref{table:all} and Figure~\ref{fig:all}.
As with CCSD(T), the linearisation barrier values from FNDMC converge to
a value close to the ZPE-corrected Raman benchmark
adopted in this study of 0.36 kcal/mol for the largest basis sets.
Also observed is the independence of the linearisation barrier
calculated from the basis set used to form the initial trial wavefunction using DFT-B3LYP 
relative to the other two calculation methods.
The influence of the different basis sets on the end result of the FNDMC calculations
directly translates into how they affect the trial wavefunction nodal surfaces.
The amplitudes of the trial wavefunction, meanwhile, do not affect the end result
at ($\tau \rightarrow \infty$),
which, in turn, limits the dependence of the end result on the basis sets used to form the trial wavefunction.

\begin{figure}[htbp]
  \centering
    \includegraphics[width=\linewidth]{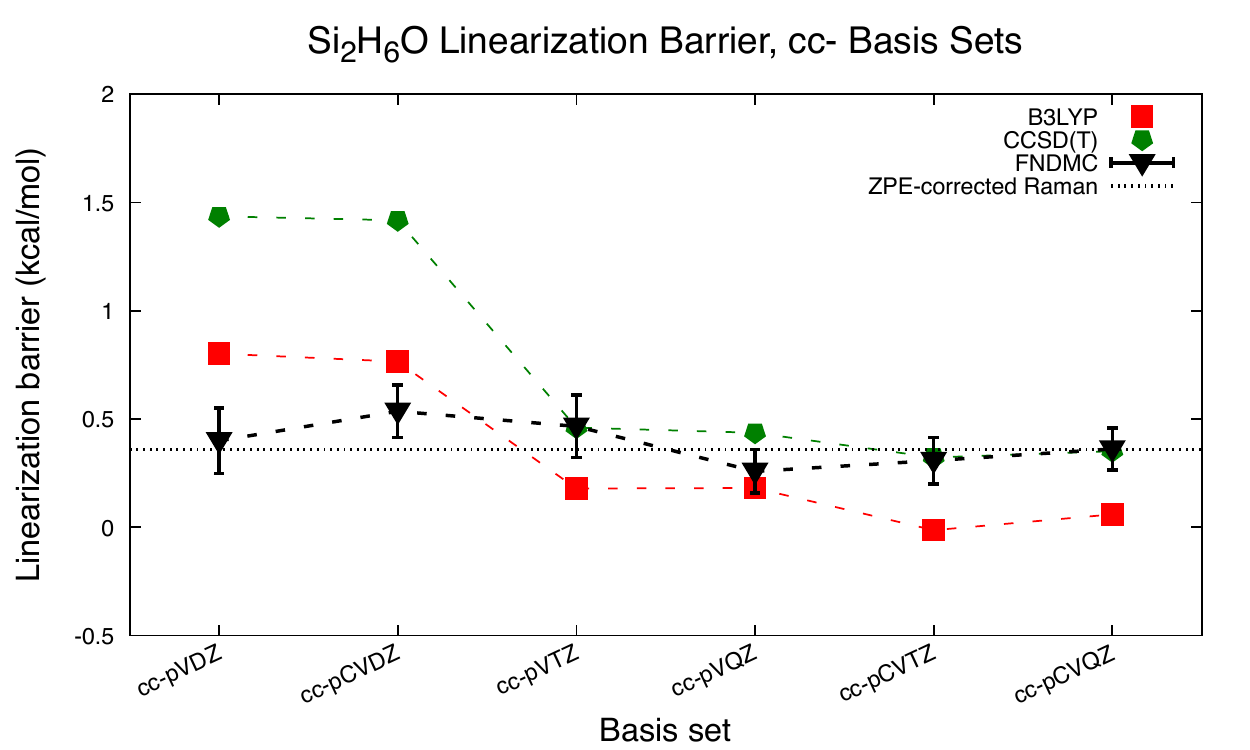}
    \caption{Linearisation barrier of disiloxane calculated using cc- basis sets, 
      sorted based on the quality of the nodal surface. 
      Dashed lines serve to better illustrate the trends present in the data.}
    \label{fig:ccsort}
\end{figure}

\vspace{2mm}
The quality of the trial wavefunction nodal surface
(how close it resembles the true ground-state wavefunction nodal surface)
is reflected in the total energy values of the FNDMC calculation
(the fixed-node variational principle) ~\cite{2001FOU}.
Therefore, the expected values of the total energy in the all-electron FNDMC calculations 
(such as those conducted in this study) are 
good indicators of the quality of the trial wavefunction nodal surfaces,
because all-electron FNDMC calculations retain the variational principle with respect to the total energy of the ground-state.
These absolute values are shown in Table~\ref{table:toten},
sorted in accord with the quality of the nodal surface
(from low quality, high total energy, to high quality, low total energy),
and displayed in Figure~\ref{fig:ccsort}.
It can be observed that the cc-pCVTZ produces a nodal surface of
slightly better quality than cc-pVQZ,
indicating that the electron correlation
between the core and valence electrons can be
a significant factor for improvement
beyond the triple zeta level of atomic orbital description.
%%%%%%%%%%%%%%%%%%%%%%%%%%%%%%%%%%%%%%%%%%%%%%%%%%%%%%%%%%%%%%%%%%%%%%%%%%%%%%%%
%%%%%%%%%%%%%%%%%%%%%%%%%%%%%%%%%%%%%%%%%%%%%%%%%%%%%%%%%%%%%%%%%%%%%%%%%%%%%%%%
%%%%%%%%%%%%%%%%%%%%%%%%%%%%%%%%%%%%%%%%%%%%%%%%%%%%%%%%%%%%%%%%%%%%%%%%%%%%%%%%
\begin{table*}[htbp]
 \centering
 \caption{Total energies from FNDMC calculations, sorted from highest to lowest.}
 \label{table:toten}
 \begin{tabular}{lcc}
  \multirow{2}{*}{Basis set} & \multicolumn{2}{c}{$E_{\rm total}$ [Hartree]}\\\cline{2-3}
  & Delinear & Linear\\
  \hline
  cc-pVDZ         & -657.7994(2) & -657.7988(2)\\
  cc-pCVDZ\phantom{000} & \phantom{0}-657.8062(1)\phantom{0} & \phantom{0}-657.8053(1)\phantom{0}\\
  cc-pVTZ         & -657.8270(2) & -657.8263(2)\\
  cc-pVQZ         & -657.8366(1) & -657.8362(1)\\
  cc-pCVTZ        & -657.8376(1) & -657.8371(1)\\
  cc-pCVQZ        & -657.8437(1) & -657.8431(1)\\
  \hline
 \end{tabular}
\end{table*}
%%%%%%%%%%%%%%%%%%%%%%%%%%%%%%%%%%%%%%%%%%%%%%%%%%%%%%%%%%%%%%%%%%%%%%%%%%%%%%%%

%%%%%%%%%%%%%%%%%%%%%%%%%%%%%%%%%%%%%%%%%%%%%%%%%%%%%%%%%%%%%%%%%%%%%%%%%%%%%%%%
\subsection{Effects of basis sets}
\label{dis:basis}
%%%%%%%%%%%%%%%%%%%%%%%%%%%%%%%%%%%%%%%%%%%%%%%%%%%%%%%%%%%%%%%%%%%%%%%%%%%%%%%%
Disiloxane linearisation barrier dependence on the basis set was observed for all cases.
In agreement with previous studies~\cite{2008DER},
this dependence is more significant than the methodologies used in the {\it ab initio} calculations,
particularly considering the double zeta-level basis sets cc-pVDZ and cc-pCVDZ.
Triple zeta basis sets seem to offer the minimum level of description to reliably
predict the linearisation barrier,
whereas quadruple zeta basis sets result in an extremely good prediction, 
especially for the CCSD(T) and FNDMC calculations.
Figure~\ref{fig:all} clearly shows very similar trends for each calculation method 
with a substantial dependence on the basis set indicated, particularly towards the smaller sets.
Convergence of the disiloxane linearisation barrier is
generally observed for all three calculation methods, 
albeit not necessarily converging to the same value.

\vspace{2mm}
Previous theoretical studies suggest that this converging trend
is attributed to the increasing addition of polarisation functions~\cite{2008DER}
within the basis sets used.
Adding polarisation functions serves to better reproduce dynamical correlations in the system.
This is in line with previous theoretical works with semi-empirical methods
implying that calculations without electron correlation favour the linear structure, 
thereby resulting in negative values of linearisation barrier.
This is also reflected in preceding geometry optimisation calculations 
without polarisation functions, resulting in Si-O-Si angles close to 170\deg.~\cite{1994CSO,1990KOP}.
Therefore, it is expected that both cc-pV$x$Z and cc-pCV$x$Z basis sets 
should converge to a reliable predicted linearisation barrier value 
because polarisation functions are systematically included with an increasing description of the atomic orbitals.

\begin{figure}[htbp]
  \centering
    \includegraphics[width=\linewidth]{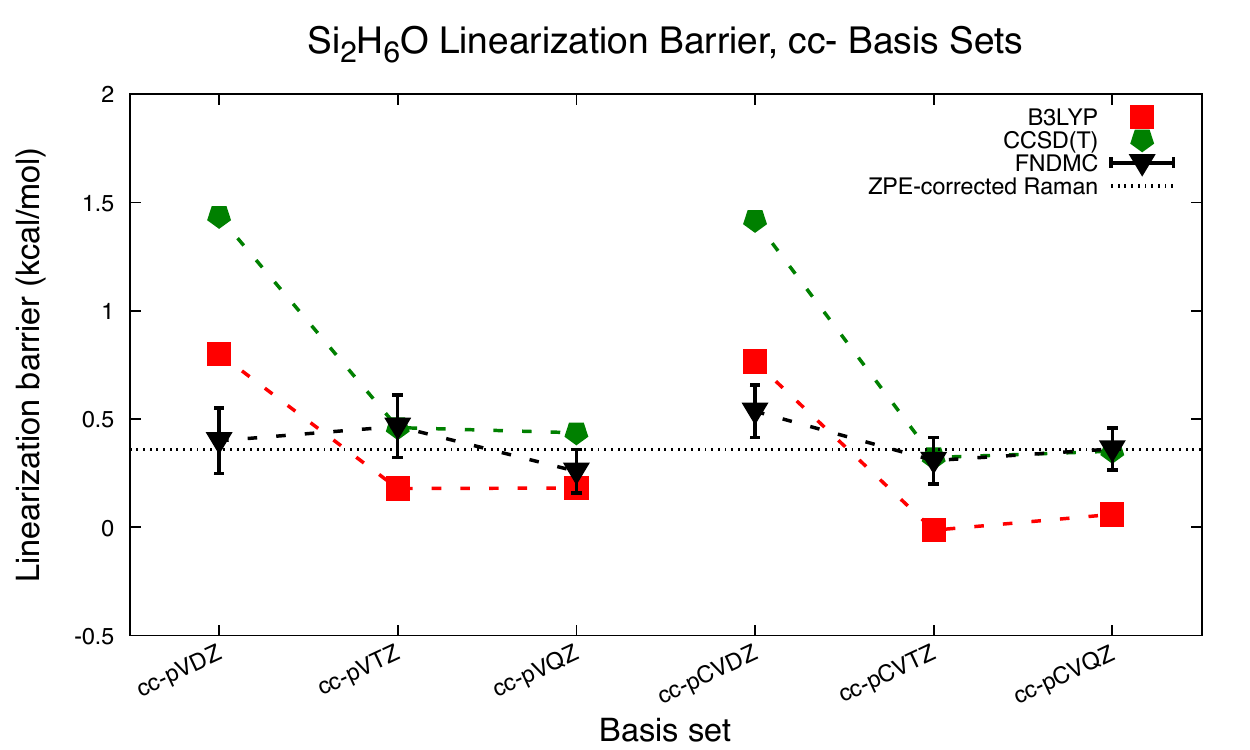}
    \caption{Linearisation barrier of disiloxane calculated using cc-basis sets, 
      separated into standard and core-valence-correlated versions (left and right). 
      Dashed lines serve to better illustrate the trends present in the data.}
    \label{fig:cc}
\end{figure}

\vspace{2mm}
From Figure~\ref{fig:cc}, a comparison can be observed
between the correlation consistent cc-pV$x$Z basis sets and its
core-valence correlated counterpart cc-pCV$x$Z.
The B3LYP and CCSD(T) linearisation barrier trends closely follow one another, 
with CCSD(T) cc-pCV$x$Z in particular 
converging to a nearly identical value as that of the ZPE-corrected Raman.
Meanwhile, the FNDMC results show a slight difference in the trends between the two families 
when observing the respective expectation values of the linearisation barriers.
It should be noted, however,
that the stochastic nature of FNDMC diminishes the significance of the energy trends 
(which need to be considered probabilistically).

\vspace{2mm}
Figure~\ref{fig:ccsort} shows that
FNDMC is overall less dependent on the basis set used
compared to both B3LYP and CCSD(T).
Although previous studies have recommended treating disiloxane with the cc-pVTZ basis set at a minimum, 
cc-pVDZ is shown to generate a sufficiently high quality nodal surface 
for use in DMC calculations, giving a linearisation barrier 
in good agreement with the experimental values.
Even for the smallest basis sets tested in this study, 
FNDMC shows relatively more accurate values of the linearisation barrier, 
and therefore less dependence on the basis set used to form the trial wavefunction per the initial expectations.
%%%%%%%%%%%%%%%%%%%%%%%%%%%%%%%%%%%%%%%%%%%%%%%%%%%%%%%%%%%%%%%%%%%%%%%%%%%%%%%%
%%%%%%%%%%%%%%%%%%%%%%%%%%%%%%%%%%%%%%%%%%%%%%%%%%%%%%%%%%%%%%%%%%%%%%%%%%%%%%%%

%%%%%%%%%%%%%%%%%%%%%%%%%%%%%%%%%%%%%%%%%%%%%%%%%%%%%%%%%%%%%%%%%%%%%%%%%%%%%%%%
\subsection{Effects of methodologies}
\label{dis:calc}
%%%%%%%%%%%%%%%%%%%%%%%%%%%%%%%%%%%%%%%%%%%%%%%%%%%%%%%%%%%%%%%%%%%%%%%%%%%%%%%%
In line with previous studies, we find the dependence on
the methods weaker than that on the basis sets.
With increasing levels of basis sets, 
DFT-B3LYP and CCSD(T) values of the linearisation barriers 
converge to approximately 0.1 and 0.3-0.4 kcal/mol, respectively.
This slight difference is in accord with 
the variance in experimental measurements~\cite{1960ARO,1977DUR,1983KOP} 
and is clearly less significant than the variance as a result of the basis sets.

\vspace{2mm}
Previous expectations on the trend in methodologies are
derived from previous studies~\cite{1990KOP,1990SHA},
particularly the study by Koput in 1990~\cite{1990KOP}, 
in which the inclusion of the electron correlation proved 
vital to predicting the energetic favourability of the non-linear structure of disiloxane because 
the SCF calculation produced a near-linear structure of disiloxane.
This and other studies ~\cite{1990SHA} gave rise to the general expectation 
that the inclusion of an electron correlation is important 
to properly model the structure of disiloxane.
As previously mentioned, it can be observed that 
the B3LYP exchange-correlation functional is insufficient 
to properly recover the electron correlation 
and thereby predict the disiloxane linearisation barrier.
Meanwhile, CCSD(T) and FNDMC calculations 
both converge towards the ZPE-corrected Raman value, 
indicating a proper accounting of the dynamical correlation.

\vspace{2mm}
FNDMC has the added advantage of being able to 
reliably predict the energetics of disiloxane with smaller basis sets.
With better scalability for application 
in high performance computing (HPC) systems 
and the availability of pseudopotentials, 
FNDMC is a promising alternative method to 
CCSD(T) for describing the Si-O-Si bonds, 
particularly for larger systems.

%%%%%%%%%%%%%%%%%%%%%%%%%%%%%%%%%%%%%%%%%%%%%%%%%%%%%%%%%%%%%%%%%%%%%%%%%%%%%%%%
\section{\label{sec:conclusion}CONCLUSION}
The Si$_2$H$_6$O linearisation barrier was calculated using three separate {\it ab initio} methods, 
DFT-B3LYP, CCSD(T), and FNDMC, 
with six different basis set choices 
in line with expectations derived from previous theoretical studies on disiloxane.
Similar with previous studies, we observed 
that the systematic inclusion of polarisation functions, 
along with an increasing level of description for atomic orbitals, 
eventually result in a reliable prediction of the disiloxane linearisation barrier.
All calculation methods eventually produced converged values 
with increasing level of basis sets for the linearisation barrier, 
with 0.1 kcal/mol for DFT-B3LYP and 0.3-0.4 kcal/mol for CCSD(T), respectively, 
in line with previous theoretical studies~\cite{2008DER}, 
and similar expectation values of the barrier for FNDMC calculations.
The agreement between the experimental measurements and the DFT-B3LYP results at 0.3 kcal/mol 
are shown to be likely accidental. 
ZPE-corrected experimental measurements are 
in good agreement with the CCSD(T) and FNDMC results, 
with the ground state linearisation barrier taken at 0.36 kcal/mol.
FNDMC is shown to be least dependent on 
the choice of basis set among the three calculation methods applied, 
in line with initial expectations owing to the nature of FNDMC.

%%%%%%%%%%%%%%%%%%%%%%%%%%%%%%%%%%%%%%%%%%%%%%%%%%%%%%%%%%%%%%%%%%%%%%%%%%%%%%%%
\section*{ACKNOWLEDGEMENTS}
The computations in this study were conducted 
using the facilities of 
Research Center for Advanced Computing 
Infrastructure at JAIST. 
R.M. would like to extend his appreciation for the financial support from 
MEXT-KAKENHI (19H04692 and 16KK0097), 
FLAGSHIP2020 (project nos. hp190169 and hp190167 at K-computer), 
the Air Force Office of Scientific Research 
(AFOSR-AOARD/FA2386-17-1-4049;FA2386-19-1-4015), 
and JSPS Bilateral Joint Projects (with India DST). 
K.H. would like to thank the HPCI System Research Project (Project ID: hp190169) and
MEXT-KAKENHI (JP16H06439, JP17K17762, JP19K05029, and JP19H05169) for the financial support.

\bibliography{references}

\end{document}